\documentclass[runningheads]{llncs}
\usepackage{graphicx}
\usepackage{subfig}
\usepackage{amsmath}
\usepackage{csvsimple}
\usepackage{xcolor}

\begin{document}
\title{Scale-Space Autoencoders for Unsupervised Anomaly Segmentation in Brain MRI}

\titlerunning{Scale-Space Autoencoders}
\author{Christoph Baur\inst{1} \and Benedikt Wiestler\inst{4} \and Shadi Albarqouni\inst{1,2} \and Nassir Navab\inst{1,3}}
%index{Baur, Christoph}
%index{Wiestler, Benedikt}
%index{Albarqouni, Shadi}
%index{Navab, Nassir}
%
\authorrunning{C. Baur et al.}

\institute{Computer Aided Medical Procedures (CAMP), TU Munich, Germany\\
\and Computer Vision Laboratory, ETH Zurich, Switzerland\\
\and Whiting School of Engineering, Johns Hopkins University, Baltimore, United States\\
\and Department of Diagnostic and Interventional Neuroradiology, Klinikum rechts der Isar, TU Munich, Germany}
\maketitle              % typeset the header of the contribution
\begin{abstract}
Brain pathologies can vary greatly in size and shape, ranging from few pixels (i.e. MS lesions) to large, space-occupying tumors. Recently proposed Autoencoder-based methods for unsupervised anomaly segmentation in brain MRI have shown promising performance, but face difficulties in modeling distributions with high fidelity, which is crucial for accurate delineation of particularly small lesions. Here, similar to these previous works, we model the distribution of healthy brain MRI to localize pathologies from erroneous reconstructions. However, to achieve improved reconstruction fidelity at higher resolutions, we learn to compress and reconstruct different frequency bands of healthy brain MRI using the laplacian pyramid. In a range of experiments comparing our method to different State-of-the-Art approaches on three different brain MR datasets with MS lesions and tumors, we show improved anomaly segmentation performance and the general capability to obtain much more crisp reconstructions of input data at native resolution. The modeling of the laplacian pyramid further enables the delineation and aggregation of lesions at multiple scales, which allows to effectively cope with different pathologies and lesion sizes using a single model.

\end{abstract}
\section{Introduction}

Supervised Deep Learning has indisputably shown great performance in the segmentation of medical images, including pathologies in brain MRI. However, these models make assumptions on the nature of pathologies they try to segment based on the labeled data they are trained from, in which rare cases might not be adequately covered and thus can potentially not be delineated properly. Generally, the unavailability of large quantities of labeled data poses a burden for the field. Recently, unsupervised representation learning and generative modeling based frameworks have emerged as promising tools to detect and segment arbitrary pathologies in MRI, without calling for pixel-precise expert annotations.

Methods based on GANs model the distribution of normal retinal OCT data and rely on the GANs' incapability to recover anomalous samples from the modeled distribution \cite{schlegl2017unsupervised,schlegl2019f}. Similarly, in the context of brain imaging, Variational Autoencoders\cite{zimmerer2019unsupervised,zimmerer2018context,pawlowski2018unsupervised} (VAEs), Adversarial Autoencoders\cite{chen2018unsupervised} (AAEs) and combinations of GANs and VAEs\cite{baur2018deep} have been proposed to model the distribution of healthy brain MRI. The feed-forward nature of these approaches allows to efficiently obtain reconstructions of input data. In those reconstructions anomalies likely have vanished as they are not part of the modeled distribution. The variational properties of these frameworks also allow to project input samples to a probabilistic latent space and to restore more likely, lesion-free counterparts by walking along the manifold\cite{you2019uad}. Although promising results have been reported, some important aspects have not yet been adequately addressed: i) different pathologies appear at different sizes and might call for different image resolutions; ii) at high resolution, reconstruction fidelity is paramount to be able to delineate small lesions with precision, but frameworks like VAEs can only provide blurry, coarse reconstructions.

Here, we propose a framework for unsupervised anomaly segmentation based on the Laplacian Pyramid, tailored around the family of Autoencoders (AEs). Our approach allows to compress and reconstruct MR images of the brain with high fidelity while successfully suppressing anomalies. More precisely, inspired by \cite{dorta2017laplacian}, we model the distribution of the scale-space representation of healthy brain MRI rather than actual image pixels. A comparison to classic AEs and other AE-based State-of-the-Art on three different datasets with different pathologies shows both superior segmentation performance and higher reconstruction fidelity. The inherent multi-scale nature of the laplacian pyramid also allows us to segment anomalies at different resolutions and to aggregate the results, which further improves the performance and gives insights into which resolution is appropriate for diseases such as MS and Glioblastoma.
\section{Methodology}

\begin{figure}
\includegraphics[width=\textwidth]{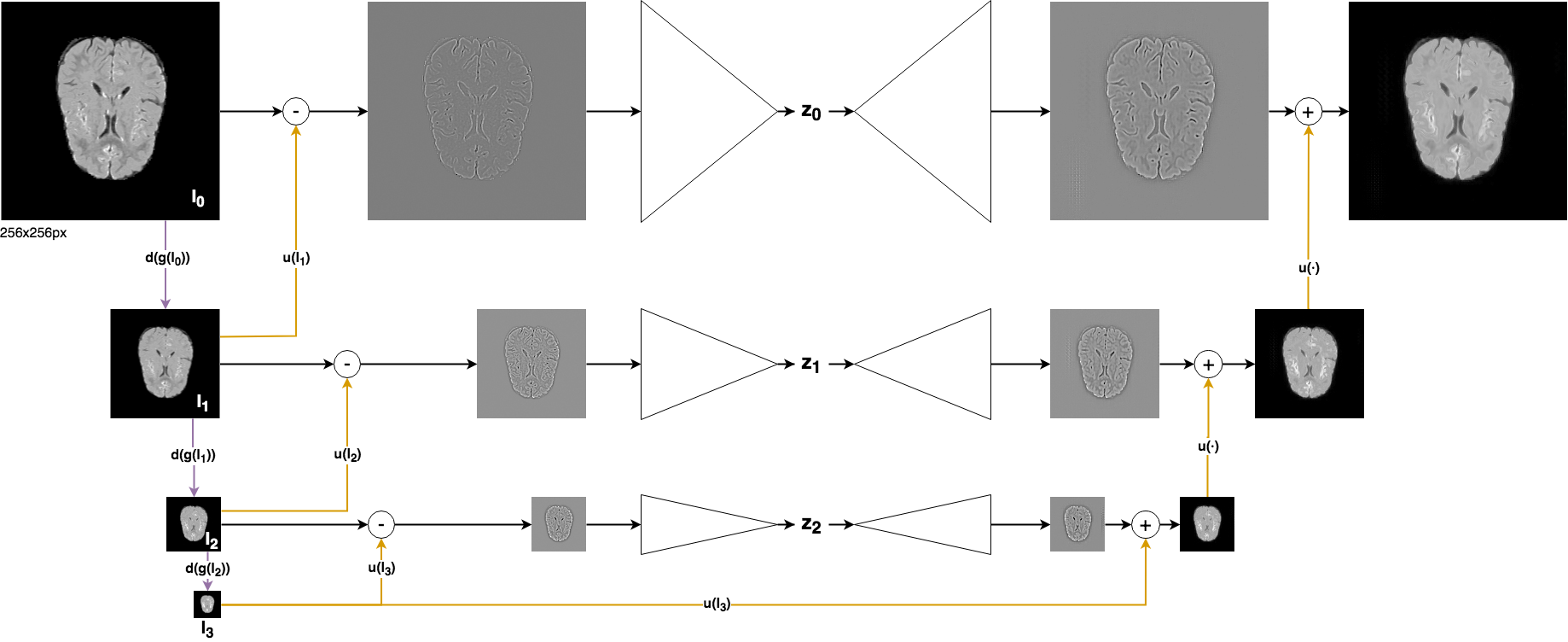}
\caption{An overview of the Scale-Space Autoencoder (SSAE) framework. A sample is decomposed into a 3-level laplacian pyramid, and every level uses a separate AE to compress and reconstruct the respective high frequency components.}
\label{fig:framework}
\end{figure}

Similar to previous work, we rely on modeling healthy anatomy with encoder-decoder networks and aim to localize anomalies from reconstruction residuals. However, we do not model the intensity distribution directly. Instead, we split the frequency band of the input data by learning to compress and reconstruct the laplacian pyramid of healthy brain MRI. 

Given a gaussian kernel $g_{\sigma}(\cdot)$ with variance $\sigma$, a downsampling operator $d(\cdot)$ and an upsampling operator $u(\cdot)$, a laplacian pyramid with $K$ levels can be obtained by repeatedly smoothing and downsampling an input image $\mathbf{x}$, i.e.

\begin{align*}
    \mathbf{I}_0 &= \mathbf{x} \\
		\mathbf{I}_k &= d(g_{\sigma}(\mathbf{I}_{k-1})) \qquad \forall 0 < k \leq K
\end{align*}

and determining the high frequency residuals $\mathbf{H}_k$ at each level $k$:

\begin{equation}
		\mathbf{H}_k = \mathbf{I}_k - u(\mathbf{I}_{k+1}) \qquad \forall 0 \leq k < K
\end{equation}

An image $\mathbf{x}$ is completely represented by the low-resolution image $\mathbf{I}_K$ after $K$ downsamplings and the high frequency residuals $\mathbf{H}_0,...\mathbf{H}_{K-1}$. A reconstruction can be obtained recursively via

\begin{equation}
\label{eq:rec}
		\mathbf{\hat{x}} = \sum_{k=0}^{K-1} u(\mathbf{I}_{K-k}) + \mathbf{H}_{K-1-k}
\end{equation}

Let $\mathcal{X}_{H}$ be a set of healthy brain MR slices and $\mathbf{x}$ be a single sample $\in \mathcal{X}_{H}$. For every level $k$ of the pyramid, we model the distribution of the respective healthy high frequency components $\mathbf{H}_k$ with an encoder-decoder network $\mathcal{M}_k(\cdot)$ by minimizing the discrepancy between $\mathbf{H}_k$ and its reconstruction $\mathbf{\hat{H}}_k = \mathcal{M}_k(\mathbf{H}_k)$ (see Fig. \ref{fig:framework}). To account for upsampling inaccuracies, we do not minimize the reconstruction error on the high frequency residuals directly. Instead, as a proxy, we minimize the difference between $\mathbf{I}_k$ and their reconstructed counterpart $\mathbf{\hat{I}}_k = u(\mathbf{\hat{I}}_{k+1}) + \mathbf{\hat{H}}_k$:

\begin{equation}
		\mathcal{L}_k = \ell_2(\mathbf{I}_k, \mathbf{\hat{I}}_k) = \ell_2(\mathbf{I}_k, u(\mathbf{\hat{I}}_{k+1}) + \mathbf{\hat{H}}_k)
\end{equation}

The overall loss is a weighted sum of losses at all scales:

\begin{equation}
		\mathcal{L} = \sum_{k=0}^K \lambda_k\mathcal{L}_k
		\label{eq:loss_total}
\end{equation}

Since the laplacian pyramid of an image is often referred to as its \emph{scale-space representation}, we refer to the resulting set of encoder-decoder networks as the Scale-Space Autoencoder (SSAE). The underlying encoder-decoder network $\mathcal{M}_k(\cdot)$ can be arbitrarily defined as a deterministic Autoencoder or as a VAE.

\subsection{Anomaly Detection}

Given a trained model and the scale-space representation of an image, it can be reconstructed at different resolutions from the recursive aggregation:

\begin{equation}
 		\mathbf{\hat{x}}_k = \mathbf{\hat{I}}_k = \sum_{i=k}^{K-1} u(\mathbf{\hat{I}}_{K-i}) + \mathcal{M}_k(\mathbf{H}_{K-1-i})
\end{equation}

Assuming that a model $\mathcal{M}_k$ is not capable to reliably reconstruct high frequency components of anomalies, an anomaly segmentation can be obtained from the residuals among $\mathbf{I}_k$ and $\mathbf{\hat{I}}_k$:

\begin{equation*}
    \mathbf{r}_k = \mathbf{I}_k - \mathbf{\hat{I}}_k
\end{equation*}

The recursive relation in Eq. \ref{eq:rec} can also be applied on the residuals $\mathbf{r}_k$ to obtain an aggregated residual image $\mathbf{r}$ at full resolution, i.e. a multi-scale aggregation of lesion segmentations:

\begin{equation}
 		\mathbf{r}_* = \sum_{k=0}^{K-1} u(\mathbf{r}_{K-k}) + \mathbf{r}_{K-k-1}
		\label{eq:multiscale_aggregation}
\end{equation}

\section{Experiments and Results}

\begin{figure}[t]
\centering
\includegraphics[width=\textwidth]{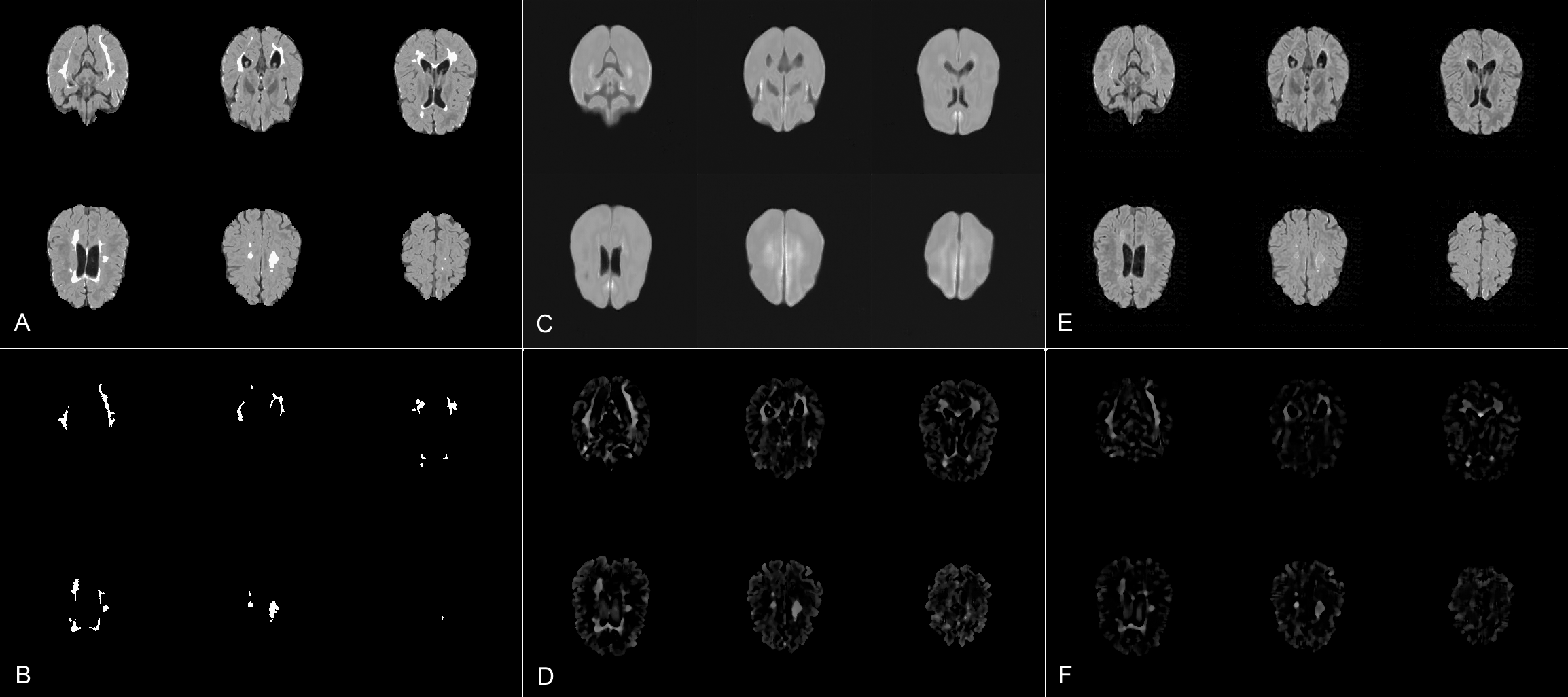}
\caption{Visual results. A: input; B: ground-truth segmentation; C: reconstruction from a normal AE; D: median-filtered residuals from C; E: reconstruction from our SSAE; F: median-filtered residuals from E. The high fidelity facilitated by our scale-space approach leads to fewer unwanted residuals.}
\label{fig:grid}
\end{figure}

In the following, we first introduce the datasets used in our experiments. In succession, we provide i) a comparison of our scale-space approach to a variety of State-of-the-Art methods, ii) a study on reconstruction fidelity and segmentation performance at multiple resolutions on different pathologies and iii) investigations of the proposed multi-scale aggregation.

\subsection{Dataset}

For evaluating our scale-space approach and the multi-scale aggregation, we employ four different datasets. To train our models, we use the FLAIR images from a dataset $\mathcal{D}_{healthy}$ of 100 healthy subjects from our clinical partners, acquired with a Philips Achieva 3T MR scanner. For testing, we use a dataset $\mathcal{D}_{MS}$ containing FLAIR scans of 49 subjects with MS, taken with the same scanner. Further, we rely on two datasets acquired with Siemens scanners: the non-public $\mathcal{D}_{GB}$, consisting of 26 subjects with Glioblastoma, and the publicly available MS dataset $\mathcal{D}_{MSLUB}$ from University Hospital of Lublijana~\cite{lesjak2018novel}. All scans were skull-stripped using ROBEX~\cite{Iglesias:2011fb}, co-registered to the SRI24 ATLAS~\cite{Rohlfing:2009dp}, and normalized by their 98th percentile into $[0;1]$. In all our experiments, we use 2D axial slices which contain brain tissue.
 
\subsection{Implementation}
All our experiments were implemented in Python with TensorFlow and carried out on a commodity GPU. Each model was trained in batches of 8 until convergence using the ADAM optimizer with a learning rate of 0.001 and an automatic early-stopping heuristic. The lagrangian multipliers $\lambda_k$ for each stage $k$ in Eq.~\ref{eq:loss_total} were used in a one-hot fashion to train every stage of the pyramid separately, starting with the lowest level $k=3$. For smoothing the images, we use a length 5 isotropic gaussian kernel with a $\sigma$ such that $> 99\%$ of the gaussian distribution are covered , and for the upsampling operator $u(\cdot)$ we adopt bilinear interpolation.

\subsection{Comparison to State-of-the-Art}

First, we compare three different variants of our scale-space approach, i.e. a dense, spatial and variational SSAE, against a variety of State-of-the-Art (SOTA) methods on all testing datasets. We measure the area under the Precision-Recall curve (AUPRC) and the optimally achievable DICE-score $\lceil$DICE$\rceil$, which constitutes a dataset-specific theoretical upper-bound to a models segmentation performance and is determined via a greedy search for the threshold $t$ which yields the highest DICE-score on a test set. Modus operandi is $128\times 128$px, as we were unable to obtain feasible results at higher resolution with all of the SOTA methods. Results are reported in Table \ref{tab:sota}. Among all reconstruction-based methods, our scale-space models always show noticeable improvements over their traditional counterpart, with the SSVAE being slightly inferior to the spatial and dense SSAE. However, on $\mathcal{D}_{MS}$ and $\mathcal{D}_{GB}$, the costly, iterative restoration-based approach from You et al.~\cite{you2019uad} shows the best overall performance.

\begin{table}[!t]
\caption{Variants of our scale-space approach compared to SOTA methods in terms of AUPRC and $\lceil$DICE$\rceil$ (higher is better). Methods marked with an * share the same model complexity. Top-2 methods in each column are bold-faced.}
\label{tab:sota}
\centering
\csvreader[tabular=|l|l|l|l|l|l|l|,
    table head= \hline & \multicolumn{2}{c|}{\textbf{$\mathcal{D}_{MS}$}} & \multicolumn{2}{c|}{\textbf{$\mathcal{D}_{GB}$}} & \multicolumn{2}{c|}{\textbf{$\mathcal{D}_{MSLUB}$}}  \\ \hline 
		\textbf{Approach} &
		\textbf{AUPRC} & \textbf{$\lceil$DICE$\rceil$}
		& \textbf{AUPRC} & \textbf{$\lceil$DICE$\rceil$}
		& \textbf{AUPRC} & \textbf{$\lceil$DICE$\rceil$}\\ \hline,
    late after line=\\\hline, separator=semicolon]%
    {csv/results.sota.top2.csv}{
    Approach=\approach,MSKRI AUPRC=\auprcms, MSKRI BPDICE=\bpdicems, MSKRI AUROC=\mskriauroc, GBKRI AUPRC=\auprcgb, GBKRI BPDICE=\bpdicegb, GBKRI AUROC=\gbkriauroc, MSLUB AUPRC=\auprcmslub, MSLUB BPDICE=\bpdicemslub, MSLUB AUROC=\mslubauroc
    }%
    {
     \approach & \auprcms & \bpdicems & \auprcgb & \bpdicegb & \auprcmslub & \bpdicemslub 
    }%
\end{table}
 
\subsection{Reconstruction Fidelity}
 
\begin{figure}[ht]
\centering
 \includegraphics[width=\textwidth]{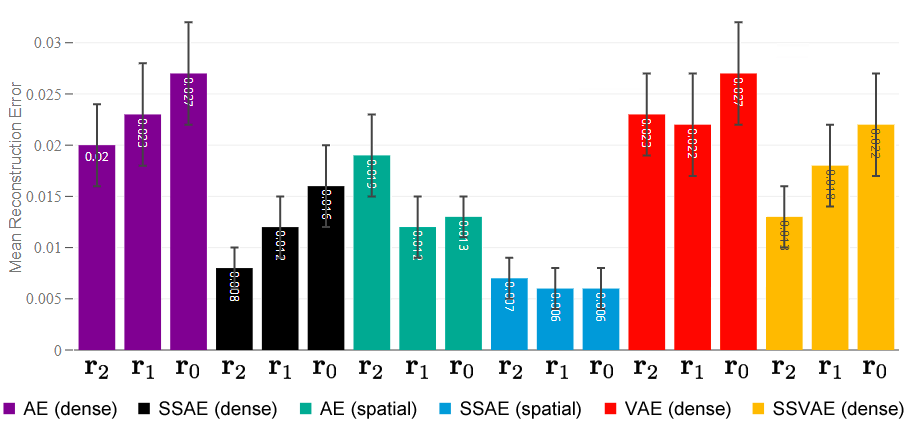}
 \caption{Normalized Reconstruction-Errors at different resolutions using different AE and SSAE models on held-out healthy validation data (lower is better).}
 \label{fig:recerr_healthy_multiscale}
\end{figure}
 
Next, we compare variants of AEs, i.e. dense AE, spatial AE and a VAE, against their scale-space counterparts in terms of their reconstruction capabilities. Again, all corresponding models share the same architecture and model complexity for a fair comparison. To measure fidelity, we collect the pixel-wise $\ell_1$-errors among all healthy validation input slices and their reconstructions, normalized by the total number of pixels. Fig.~\ref{fig:recerr_healthy_multiscale} shows the corresponding statistics on $\mathbf{r}_0 = 256\times256$px, $\mathbf{r}_1 = 128\times128$px and $\mathbf{r}_2 = 64\times64$px. The upper limit of $256\times 256$px was set by our training data $\mathcal{D}_{healthy}$. In comparison to their AE counterpart, all scale-space models show substantially lower reconstruction errors at all scales. As expected, reconstruction errors increase with image resolution, as the modeling task becomes more complex. The lowest error is achieved by a spatial SSAE, which reconstructs data almost perfectly due to the low level of compression in its bottleneck. Interestingly, a dense SSAE is on par with a spatial AE, although it loses any spatial cues in its latent space. The achieved high fidelity can also be seen in our visual results (Fig. \ref{fig:grid}). 

\subsection{Investigating Resolution and Multi-scale Aggregation}

\begin{table*}[!t]
\caption{Segmentation comparing dense, spatial AEs and variational AEs/SSAEs at different resolution as well as our multi-scale aggregation.}
\label{tab:maggr}
\centering
\csvreader[tabular=|l|l|l|l|l|l|l|l|,
    table head= \hline & &  \multicolumn{2}{c|}{\textbf{$\mathcal{D}_{MS}$}} & \multicolumn{2}{c|}{\textbf{$\mathcal{D}_{GB}$}} & \multicolumn{2}{c|}{\textbf{$\mathcal{D}_{MSLUB}$}}  \\ \hline 
		\textbf{Approach} &
		\textbf{Resolution} &
		\textbf{AUPRC} & \textbf{$\lceil$DICE$\rceil$}
		& \textbf{AUPRC} & \textbf{$\lceil$DICE$\rceil$}
		& \textbf{AUPRC} & \textbf{$\lceil$DICE$\rceil$}\\ \hline,
    late after line=\\\hline]%
    {csv/results.selected.csv}{Approach=\approach, Stage=\stage,  
    MSKRI AUROC=\aurocms, MSKRI AUPRC=\auprcms, MSKRI BPDICE=\bpdicems, MSKRI DICE=\dicems, MSKRI Rec.-Error=\recerrms,
    GBKRI AUROC=\aurocgb, GBKRI AUPRC=\auprcgb, GBKRI BPDICE=\bpdicegb, GBKRI DICE=\dicegb, GBKRI Rec.-Error=\recerrgb,
    MSLUB AUROC=\aurocmslub, MSLUB AUPRC=\auprcmslub, MSLUB BPDICE=\bpdicemslub, MSLUB DICE=\dicemslub, MSLUB Rec.-Error=\recerrmslub
    }
    {
     \approach & $\stage$ & \auprcms & \bpdicems & \auprcgb & \bpdicegb & \auprcmslub & \bpdicemslub 
    }%
\end{table*}

Finally, we compare the different scale-space and traditional AE variants by their segmentation performance on the three datasets, again measured using the AUPRC \& $\lceil$DICE$\rceil$, at different resolutions and investigate the benefits of the proposed multi-scale aggregation of residuals (Eq. \ref{eq:multiscale_aggregation}) at highest resolution (see Table \ref{tab:maggr}). For MS lesions in $\mathcal{D}_{MS}$, which has been acquired with the same scanner as our healthy training data, best AUPRC is achieved by a dense SSAE at native resolution, yielding an absolute improvement of $19\%$ over its corresponding dense AE. On $\mathcal{D}_{MSLUB}$, performance is significantly lower across the board due to lower contrast, but the dense SSAE still shows the best performance. On both datasets, additional $4\%$ can be gained by aggregating residuals from multiple scales. In contrast to MS lesions, segmentation of tumors in $\mathcal{D}_{GB}$ works best at $128\times 128$px with the majority of methods, and the proposed multi-scale aggregation shows no gains. The winning approach in this context is the spatial SSAE.

\subsection{Discussion}

The proposed scale-space formulation appears to be especially beneficial at native resolution, where it leads to considerably better reconstructions across all datasets. This is especially useful for segmenting MS lesions, which can become very small. In this context, multi-scale aggregation also turns out to be beneficial, as these lesions can vary greatly in shape and size. For large, space-occupying lesions such as Glioblastoma ($\mathcal{D}_{GB}$), a resolution of $128\times 128$px turns out to be preferable. In this context, we also find our scale-space approach not to provide much benefits, as it generates undesirably good reconstructions of large, homogenous lesions. Overall, the multi-scale aggregation leads to improvements in most of the cases, but generally is of greater value for normal AEs, whose anomaly detections appear to be more orthogonal among different resolutions and aggregate to a better consensus. Anomaly segmentations obtained from our scale-space models seem to correlate more across different resolutions.
\section{Conclusion}

In conclusion, we proposed to model normal brain anatomy in a laplacian pyramid representation to obtain high fidelity reconstructions and improved segmentation performance. We successfully demonstrate the use of this scale-space approach for unsupervised anomaly segmentation in brain MRI on different datasets with different pathologies. From the inherent multi-scale nature of our scale-space formulation, we derived a multi-scale residual aggregation technique for building an anomaly segmentation consensus among multiple resolutions, which i) turned out to be beneficial in most of the examined scenarios and ii) works for normal AEs as well. In future work, the design of a shared latent space between the different encoder-decoder networks could be investigated, and restoration approaches like \cite{you2019uad} could be adapted for our framework. Using a scale-space representation of the MR data, we also see opportunities towards improved domain invariance in unsupervised anomaly segmentation methods.
\bibliographystyle{splncs04}
\bibliography{paper}

\end{document}